\documentclass[aps,pre,amsmath,amssymb,superscriptaddress,floatfix,preprint]{revtex4}     
\usepackage{graphicx}
\usepackage{placeins}

\begin{document}
\title{Simulations of a lattice model of two-headed linear amphiphiles:  influence of amphiphile asymmetry}

\author{Douglas R. Jackson}
\affiliation{Department of Chemistry, Dalhousie University, Halifax, NS, B3H 4J3, Canada}

\author{Amir Mohareb}
\affiliation{Department of Physics, St. Francis Xavier University, Antigonish, NS, B2G 2W5, Canada}

\author{Jennifer MacNeil}
\affiliation{Department of Physics, St. Francis Xavier University, Antigonish, NS, B2G 2W5, Canada}

\author{M. Shajahan G. Razul}
\affiliation{Department of Physics, St. Francis Xavier University, Antigonish, NS, B2G 2W5, Canada}

\author{D. Gerrard Marangoni}
\affiliation{Department of Chemistry, St. Francis Xavier University, Antigonish, NS, B2G 2W5, Canada}

\author{Peter H. Poole}
\affiliation{Department of Physics, St. Francis Xavier University, Antigonish, NS, B2G 2W5, Canada}

\begin{abstract}
Using a 2D lattice model, we conduct Monte Carlo simulations of micellar aggregation of linear-chain amphiphiles having two solvophilic head groups.  In the context of this simple model, we quantify how the amphiphile architecture influences the critical micelle concentration (CMC), with a particular focus on the role of the asymmetry of the amphiphile structure.  Accordingly, we study all possible arrangements of the head groups along amphiphile chains of fixed length $N=12$ and 16 molecular units.  This set of idealized amphiphile architectures approximates many cases of symmetric and asymmetric gemini surfactants, double-headed surfactants and boloform surfactants.  
Consistent with earlier results, we find that the number of spacer units $s$ separating the heads has a significant influence on the CMC, with the CMC increasing with $s$ for $s<N/2$.  In comparison, the influence of the asymmetry of the chain architecture on the CMC is much weaker, as is also found experimentally.  
%In addition, our results highlight the distinctive properties of boloform surfactants, in comparison to other architectures.
\end{abstract}

\date{\today}
%\pacs{82.70.Uv, 07.05.Tp}
\maketitle

\section{Introduction}

Amphiphile molecules having two solvophilic head groups and the overall topology of a linear chain (or at least approximating that of a linear chain) represent an important class of surfactant.  Most prominently, this class includes the 
symmetric gemini surfactants, in which two conventional single-headed amphiphiles are joined at or near the head groups by a linear spacer~\cite{
Menger:2000p7148,
Zana:2002p6378}.  
Other examples of two-headed quasi-linear amphiphiles are certain species of double-headed~\cite{Comeau:1995p7151} and boloform surfactants~\cite{Yiv:1976p7141,Ikeda:1989p6383}.

The relationship between the architecture of amphiphilic molecules and their bulk properties in solution has long been a central theme in the study of surfactants.  This is certainly true for gemini surfactants, for which numerous studies (reviewed in Ref.~\cite{Zana:2002p6378}) have demonstrated the influence of the length of the solvophobic tails, head group size, and the length of the spacer, in controlling important properties of the solution, for example, the critical micelle concentration (CMC), and the typical size of the micelles formed (often referred to as the ``aggregation number").  

Although much of the experimental work on two-headed amphiphiles has focussed on the symmetric gemini surfactants, there are of course a large number of distinct two-headed linear architectures that are asymmetric.  Indeed, over the last decade there has been an emerging interest in the properties of ``dissymmetric gemini surfactants'' in which the two solvophobic tails are of unequal length~\cite{
Oda:1997p6508,
Huc:1999p6511,
Oda:1999p6510,
Sikiric:2002p6512,
Bai:2002p6476,
Sikiric:2003p7147,
Wang:2003p6498,
Romero:2004p6518,
Jiang:2005p6517,
Sikiric:2005p6375,
Fan:2007p6514,
Nowicki:2010p7145,
Xu:2010p6519}.  
In particular, Thomas and coworkers~\cite{Bai:2002p6476,Jiang:2005p6517} found that the CMC decreases by approximately 35\% as the asymmetry of the amphiphile increases.  However, the magnitude of this decrease is small compared to the effect on the CMC of other structural factors of gemini surfactants such as head spacing, which can change the CMC by an order of magnitude or more~\cite{
Zana:1991p6462,
Zana:2002p6378}. 
At the same time, changes in micelle size and morphology have been reported for dissymmetric gemini surfactants as a function of asymmetry~\cite{
Oda:1997p6508,
Sikiric:2002p6512,
Wang:2003p6498}.  
It would be useful to quantify the degree to which the purely geometrical influence of asymmetry is responsible for these changes, or if their origins are more subtle.

Several simulation studies have been carried out on specific cases of two-headed linear amphiphiles~\cite{
Karaborni:1994p7159,
Maiti:1998p6401,
Maiti:1998p7152,
Layn:1998p7155,
Maiti:2000p7153,
Oda:2002p7156,
Khurana:2006p7158,
Burrows:2007p6516,
Samanta:2009p7154}.
These studies have confirmed the dominant influence of tail length and head spacing on properties of the solution~\cite{Maiti:1998p6401,Maiti:1998p7152,Khurana:2006p7158}.
However, to our knowledge no simulation studies have focussed specifically on the question of the extent to which the asymmetry of the amphiphile architecture influences e.g. the CMC.  This is somewhat surprising given that symmetry, or lack thereof, is a fundamental property of any molecule, and is often an important one for determining bulk properties, e.g. the structure and properties of crystalline phases.

In this work, we present computer simulation results that examine the properties of every member of a family of two-headed linear amphiphiles for a given fixed $N$, in the context of a simple 2D lattice model.  We idealize a two-headed linear amphiphile as a flexible, linear chain made up of $N$ units.  $N-2$ of these units are solvophobic tail units, while the remaining two are solvophilic head units.  These head units can be separately located at any point along the chain.  Viewed in this purely geometric way, dozens of distinct architectures of such amphiphiles are possible even for modest values of $N\leq16$.  At the same time, each amphiphile architecture (for fixed $N$) can be precisely specified in terms of only two independent parameters, e.g. the positions along the chain of the two heads.  From the standpoint of theory and computer modelling, this family of amphiphiles thus provides an interesting case for studying the influence of structural parameters on solution properties, in the sense that it is rich enough to be interesting, but simple enough to be tractable.

Our goals are two-fold: (i) We wish to characterize a complete set of amphiphile structures in order to assess in a comprehensive way the influence of architectural parameters on the behavior of the resulting solutions.  (ii) In particular, we wish to quantify the influence of asymmetry on system properties, relative to other architectural parameters.  To achieve these goals, we use a simple 2D lattice model of an amphiphile solution, of the type introduced by Larson~\cite{
Larson:1985p6394,
Larson:1988p6393,
Larson:1989p6397,
Larson:1992p6395,
Zaldivar:2003p6348},
and sample its equilibrium properties using Monte Carlo dynamics.  Simulation models of this type have been shown to reproduce a wide range of qualitative behavior observed for surfactants solutions, and continue to play an important role in the exploration of these systems~\cite{
Care:1987p5246,
Care:1987p5248,
Brindle:1992p5247,
Stauffer:1994p6396,
Bernardes:1994p5316,
Bernardes:1996p5253,
Maiti:1998p6401,
Maiti:1998p7152,
Bhattacharya:2000p6346,
Maiti:2000p7153,
Bhattacharya:2001p5254,
Panagiotopoulos:2002p6025,
Kapila:2002p5608,
Lisal:2002p6028,
Firetto:2006p5974,
Maycock:2006p5252,
Davis:2008p5959,
Davis:2009p5950,
Davis:2009p5949,
PoorgholamiBejarpasi:2010p7140}.  
To our knowledge, an exhaustive examination via simulations of all possible amphiphile architectures of a given class, on the scale presented here, has not been carried out to date.  As shown below, our use of a simple 2D lattice model allows such a survey to be completed on a reasonable computational time scale.  While our results will necessarily be qualitative in nature, a simple lattice model allows us to focus on effects that are purely geometrical in origin, including those due to amphiphile asymmetry.

\section{Methods}

\subsection{Model}

Our model consists of $n_s$ linear amphiphile molecules on a two-dimensional 
$L \times L$ square lattice.  Unless otherwise noted, all data presented are for $L=200$.  Each amphiphile is represented as a linear chain of $N$ connected sites.  In this work, we present results for $N=12$ and $16$.  Each site along an amphiphile is either solvophobic (labelled T for ``tail") or solvophilic (labelled H for ``head").  Note that any spacer units occurring between the two heads are considered to be equivalent to tail units.  All other sites of the lattice are occupied by solvent molecules (labelled S).  The total energy of the system is given by,
\begin{equation}
{\cal H} = \sum_{\alpha \beta}\epsilon_{\alpha\beta} n_{\alpha\beta},
\label{ham}
\end{equation}
where the sum is taken over all possible nearest-neighbor pairs $(\alpha, \beta)$ of the species types T, H and S.  $\epsilon_{\alpha\beta}$ is the interaction energy of two nearest-neighbor sites on the lattice occupied by species $\alpha$ and $\beta$, and $n_{\alpha\beta}$ is the number of nearest-neighbor $(\alpha, \beta)$ contacts occurring in the system.  

Numerous simulations of micelle-forming systems have been carried out using a Hamiltonian of the form of Eq.~(\ref{ham})~\cite{
Care:1987p5246,
Care:1987p5248,
Brindle:1992p5247,
Stauffer:1994p6396,
Bernardes:1994p5316,
Bernardes:1996p5253,
Maiti:1998p6401,
Maiti:1998p7152,
Bhattacharya:2000p6346,
Maiti:2000p7153,
Bhattacharya:2001p5254,
Panagiotopoulos:2002p6025,
Kapila:2002p5608,
Lisal:2002p6028,
Firetto:2006p5974,
Maycock:2006p5252,
Davis:2008p5959,
Davis:2009p5950,
Davis:2009p5949,
PoorgholamiBejarpasi:2010p7140}.  
A wide range of choices for the interaction parameters have been shown to give systems exhibiting stable micelles.  In this work, we set $\epsilon_{\rm TS}=+1$, $\epsilon_{\rm TH}=+1$, $\epsilon_{\rm HS}=-1$ and $\epsilon_{\rm HH}=+2$, with all other interaction energies set to zero.  Our choice is similar to that of Kapila, et al.~\cite{Kapila:2002p5608}, who chose parameters equivalent to $\epsilon_{\rm TS}=+1$, $\epsilon_{\rm TH}=+1$, $\epsilon_{\rm HW}=-5.77$ and $\epsilon_{\rm HH}=+5.77$, with all others zero.  As in the present work, the aim of Ref.~\cite{Kapila:2002p5608} was to evaluate the properties of several amphiphile architectures in 2D on a square lattice, and having chain lengths similar to ours; they studied N=13 and 19, while we study N=12 and 16.  In Section II.C, we present several tests of our model to confirm that it is appropriate for our purposes.

In the following, we define the amphiphile concentration as $X=n_s/(L^2-n_s N)$, 
that is, the ratio of the number of amphiphile molecules to the number of solvent molecules.  Clusters of amphiphiles are defined as contiguous groups of adjacent amphiphiles that are connected by nearest-neighbor contacts involving either chain element (i.e. H or T units).  ``Free monomers" are isolated amphiphiles that are completely surrounded by solvent.  The free monomer concentration is defined as $X_1=n_1/(L^2-n_s N)$, where $n_1$ is the number of free monomers.

\begin{figure}\bigskip
\centerline{\includegraphics[scale=0.42]{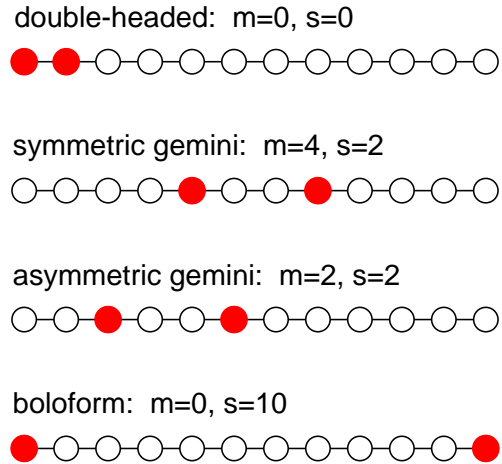}}
\caption{Four characteristic examples of amphiphile architectures for linear chains of $N=12$ having two solvophilic head units (filled circles), with all other units being solvophobic tail units (open circles).  For each architecture the corresponding $m$ and $s$ values are given.
}\label{examples}\end{figure}

In this work we examine all distinct amphiphile architectures for linear chains having two head units.  The architecture of two-headed linear amphiphiles is often specified as a triplet of integers: $m$-$s$-$k$.  Here $m$ is the number of tail units between one end of the chain and the first head unit; $s$ is the number of spacer units occurring between the two head units; and $k$ is the number of tail units between the second head unit and the other end of the chain.  For fixed $N$, only two of these integers are independent, since $N=2+m+s+k$.  Hence we specify each architecture by $m$ and $s$.  Some example architectures, and their corresponding $(m,s)$ values, are shown in Fig.~\ref{examples} for $N=12$.  Most architectures can be specified by two combinations of $(m,s)$; e.g. both $(m=0,s=0)$ and $(m=10,s=0)$ correspond to the same double-headed amphiphile for $N=12$.  In such cases, we only consider the $(m,s)$ pair having the lower value of $m$.  If $N$ is even, there are $M=\sum_{i=1}^{N/2} (2i-1)$ distinct architectures of amphiphiles containing $N$ units.  For $N=12$, $M=36$; for $N=16$, $M=64$.  Fig.~\ref{examples} illustrates that the set of these amphiphiles contains double-headed amphiphiles ($m=s=0$), symmetric gemini surfactants ($m=k$, with $m>0$ and $k>0$), 
asymmetric gemini surfactants ($m\neq k$, with $m>0$ and $k>0$), 
as well as boloform amphiphiles ($m=k=0$).  

\subsection{Simulation protocol}

We sample the equilibrium configurations of the system in the canonical ensemble using a Monte Carlo (MC) dynamics in which both amphiphile reptation and translation moves are attempted~\cite{Kapila:2002p5608,Firetto:2006p5974}.  In the reptation move, an attempt is made to move one end of the chain onto a solvent-occupied site, with the rest of the amphiphile following in train, and the displaced solvent unit moving to the site vacated by the other end of the chain.  Reptation serves to both relax the chain shape and diffuse chains through the lattice.  The translation move, in which an attempt is made to move a chain to a randomly chosen location, accelerates the diffusion of chains throughout the system volume, especially at low concentration.

We use the following procedure to equilibrate the system at a given $X$ and temperature $T$.  First, a starting configuration is generated by distributing the required number of amphiphiles at random throughout the lattice.  The initial configuration of each amphiphile is a straight line-segment oriented along the $y$-axis.  The system is then evolved using only reptation attempts until $n_s N^2$ or $30\,000$ moves are accepted, whichever is greater.  Reptation causes a single site on a surfactant molecule to execute a random walk along the path length explored by the molecule.  After $N^2$ accepted reptation moves, a given molecule will have on average moved a distance $N$ along its path length, which is sufficient to achieve a preliminary relaxation of its shape from the initially straight starting configuration.  Consequently, after $n_s N^2$ accepted reptation moves, every molecule in the system will have, on average, relaxed in shape. 

We then continue the MC trajectory by choosing reptation and translation attempts with equal probability.  The run is equilibrated for as many MC steps as are required to accumulate $n_s (10N)^2$ accepted reptation moves.  This criterion ensures that regardless of how many translation attempts are made, a sufficient number of reptation attempts have been accepted for each molecule to have (on average) diffused a path length equal to ten times its own length $N$.  

After this period of equilibration is over, an identical production phase is carried out, again selecting translation and reptation attempts with equal probability, and for as many MC steps as are required to accumulate $n_s (10N)^2$ accepted reptation moves.  All averages reported here are accumulated over this production phase.  
In addition, unless noted otherwise, all runs have been carried out for three independent starting configurations, and the results averaged.  Also, unless otherwise noted, we have evaluated the system properties for concentrations from $X=0.0005$ to $X=0.02$ in steps of $\Delta X=0.0005$.

\subsection{Properties of the model}

We have chosen our model parameters to allow for the observation of micelle-like aggregation over a range of chain lengths and architectures.  In this section, we illustrate this behavior for two distinct test cases.  The first case (denoted in the following as ``H$_1$T$_6$") models an amphiphile consisting of a single H unit attached to one end of a chain of six T units;  this corresponds to a relatively simple, single-headed surfactant.  The second case (denoted ``6-2-6") models a symmetric gemini surfactant with $N=16$ and architecture ($m$-$s$-$k$)=(6-2-6), that is, two T units separating two H units, and two symmetric tails each of six T units.  Note that the 6-2-6 amphiphile is equivalent to two H$_1$T$_6$ chains joined at the heads by two T units.

\begin{figure}\bigskip
\centerline{\includegraphics[scale=0.37]{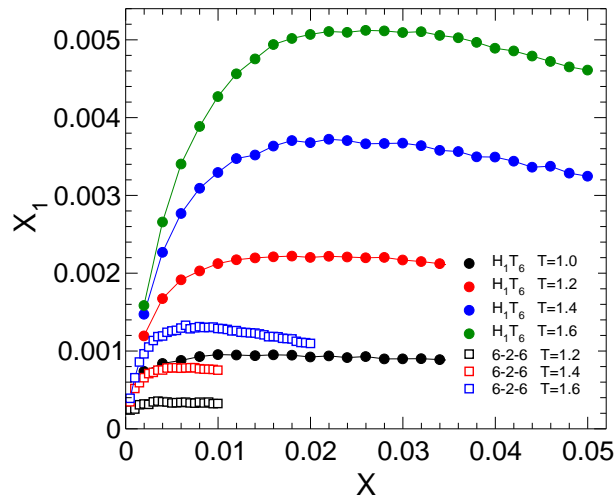}}
\caption{$X_1$ versus $X$ for the H$_1$T$_6$ and 6-2-6 systems, at several $T$.
}
\label{c1}
\end{figure}

Fig.~\ref{c1} shows the dependence of $X_1$ on $X$ for both the H$_1$T$_6$ and 6-2-6 cases, at several different $T$.  These curves display the shape consistent with a micelle-forming system, in that $X_1$ increases linearly with $X$ for small $X$, but saturates to a nearly constant value for larger $X$~\cite{
Bhattacharya:2001p5254,
Zaldivar:2003p6348}.  
The region of this crossover in the behavior of $X_1$ approximately coincides with the onset of amphiphile aggregation in the system.  The $T$ dependence of these curves, in which the saturation value of $X_1$ increases with $T$, is consistent with the behavior observed in numerous other simulations of amphiphile systems (see e.g. Ref.~\cite{Bernardes:1994p5316}).

\begin{figure}\bigskip
\centerline{\includegraphics[scale=0.37]{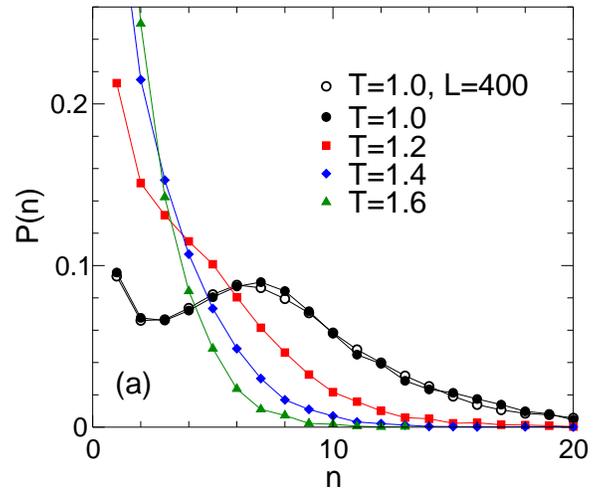}}
\bigskip\bigskip
\centerline{\includegraphics[scale=0.37]{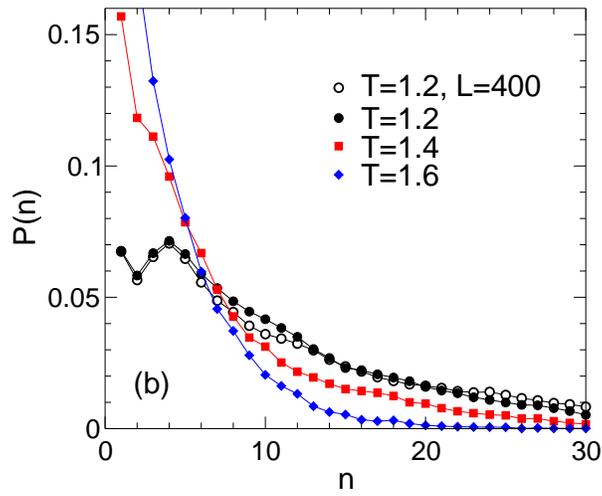}}
\caption{$P(n)$ for the (a) H$_1$T$_6$ and (b) 6-2-6 systems, at several $T$.  All curves are for $L=200$ except as indicated.  In (a) $X=0.01$; in (b) $X=0.005$.  
}\label{c2}\end{figure}

\begin{figure}\bigskip
\centerline{\includegraphics[scale=0.4]{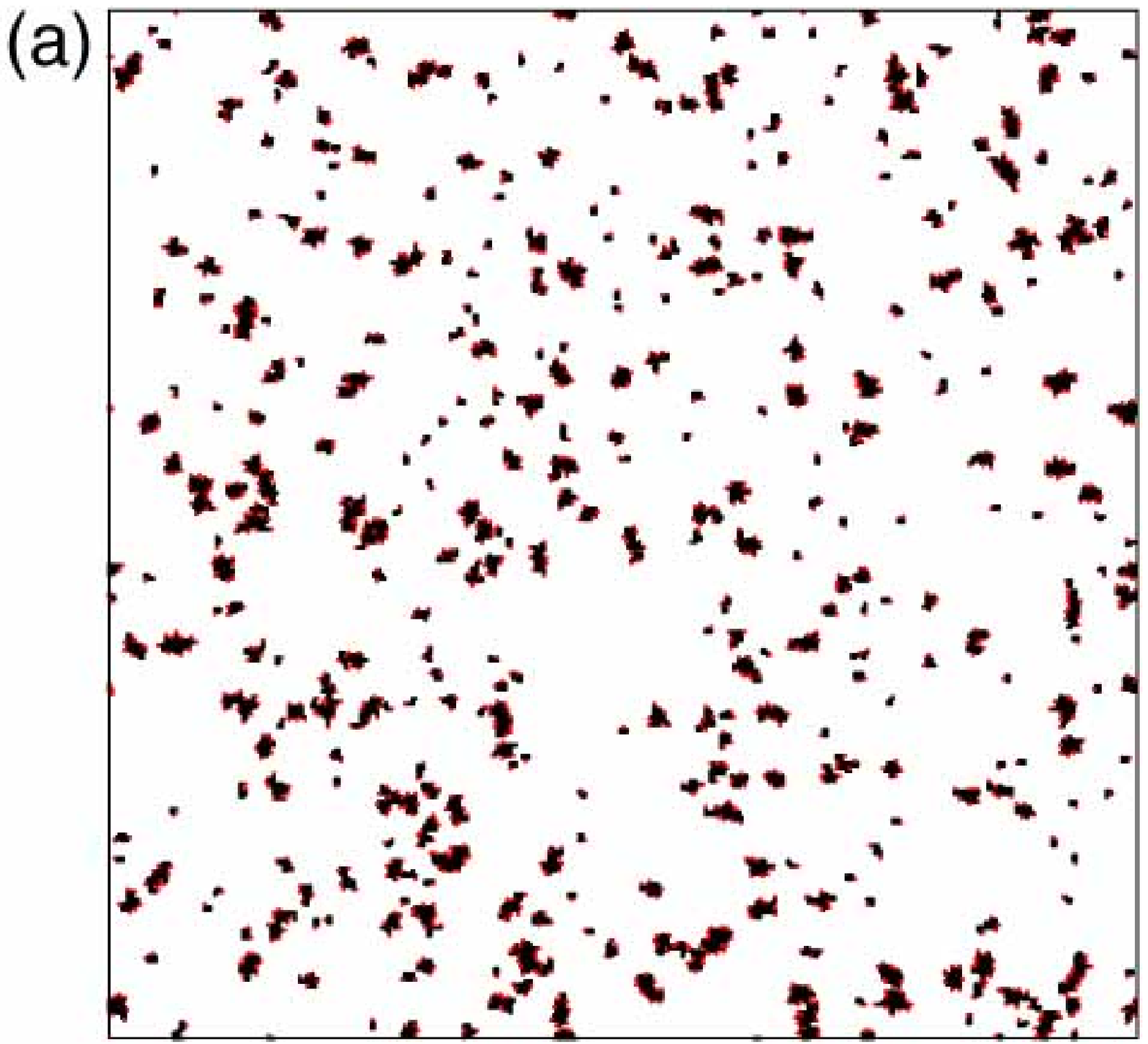}}
\bigskip\bigskip
\centerline{\includegraphics[scale=0.4]{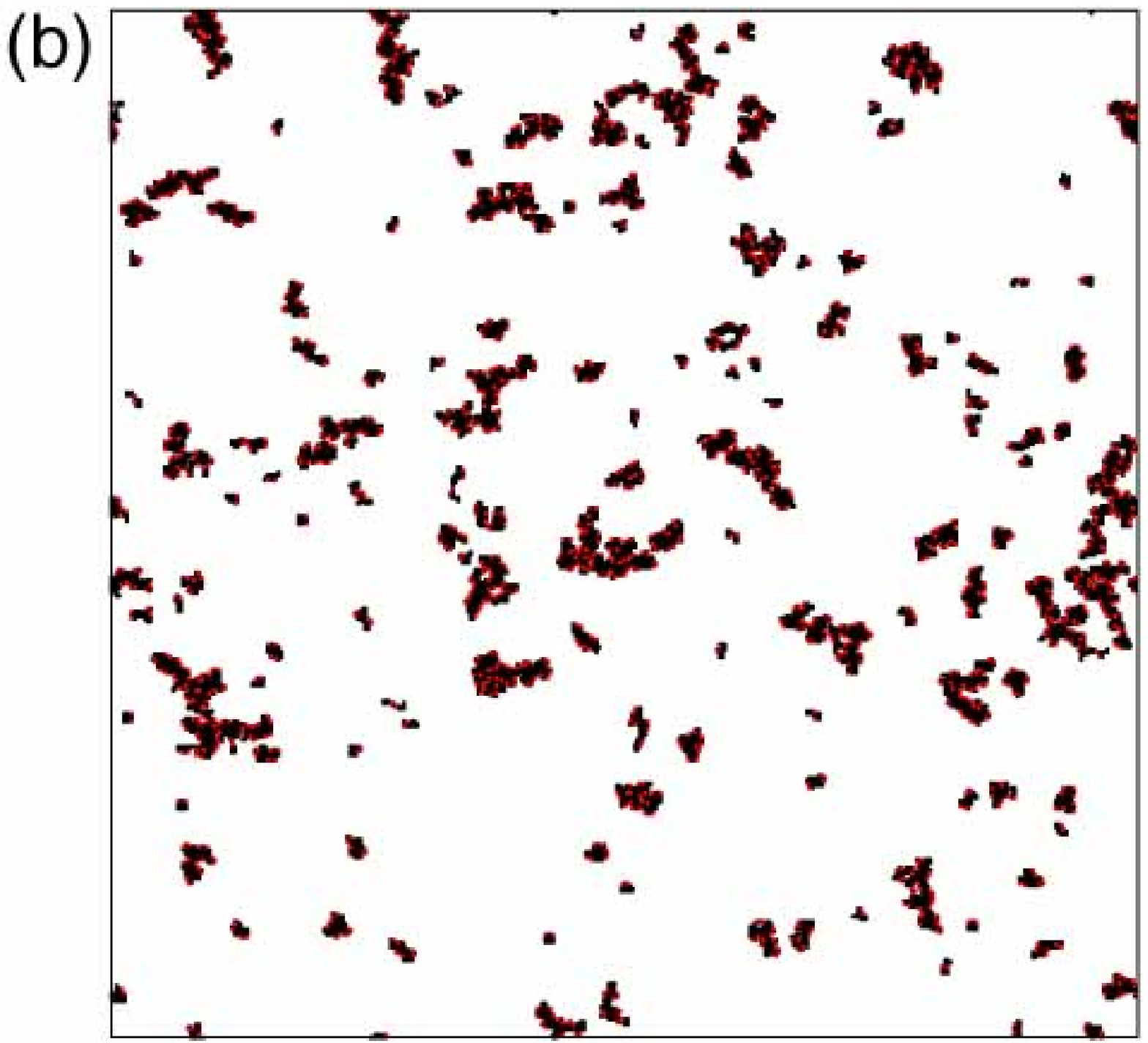}}
\caption{Snapshots of the (a) H$_1$T$_6$ and (b) 6-2-6 systems, both for $L=400$.  In (a) $T=1.0$ and $X=0.01$; in (b) $T=1.2$ and $X=0.005$.  Tail units are rendered as black line segments, and head units are red dots.
}\label{pic}\end{figure}

Aggregation of the amphiphiles is also reflected in the behavior of $P(n)$, the probability that a randomly chosen amphiphile is part of a cluster of amphiphiles of size $n$~\cite{
Care:1987p5246,
Bernardes:1994p5316,
Davis:2009p5950}.  
We compute this probability as $P(n)=nC(n)/n_s$, where $C(n)$ is the average number of clusters of size $n$.  As shown in Fig.~\ref{c2}, the emergence as $T$ decreases of a shoulder or a peak in $P(n)$ at non-zero $n$ is an indication that the aggregation process is generating clusters of a defined size.  Snapshots of both the H$_1$T$_6$ and 6-2-6 systems are shown in Fig.~\ref{pic} for the same $T$ at which a peak is observed in $P(n)$ in Fig.~\ref{c2}.

In simulations of amphiphile systems it is important to distinguish between conditions in which aggregation occurs due to stable micelle formation, and aggregation that occurs as a result of the onset of macroscopic phase separation of the amphiphiles from the solvent~\cite{Panagiotopoulos:2002p6025}.  We therefore conduct several tests to confirm the validity of the model and our simulation algorithm, as well as to ensure that a regime of micelle-like aggregation is observed:

(i) We have tested our simulation algorithm and equilibration protocol by reproducing results from a number of previous works.  In particular, by appropriate variation of the model parameters, we have reproduced the results for $X_1$ and $P(n)$ described in Refs.~\cite{Care:1987p5246,Bhattacharya:2001p5254,Maycock:2006p5252}.

\begin{figure}\bigskip
\centerline{\includegraphics[scale=0.37]{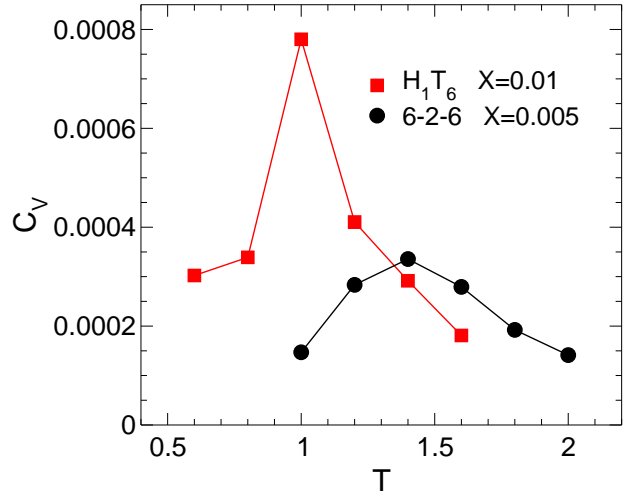}}
\caption{$C_V$ as a function of $T$ for the H$_1$T$_6$ and 6-2-6 systems.
}\label{cv}\end{figure}

(ii) In addition to the behavior of $P(n)$ shown in Fig.~\ref{c2}, the behavior of the constant-volume specific heat $C_V$, shown in Fig.~\ref{cv}, is also consistent with the formation of micelle-like aggregates.  As shown in Ref.~\cite{Bhattacharya:2001p5254}, the onset of the formation of micelles coincides with a maximum in the $T$ dependence of $C_V$.  For both the H$_1$T$_6$ and 6-2-6 systems, we find that the $T$ at which a maximum at non-zero $n$ appears in $P(n)$ corresponds to entering the $T$ regime near and below the peak in $C_V$.

(iii) In all cases reported here, we confirm that the run-time criteria described in the previous section yield stationary time series during the production phase for the system energy and $X_1$.  We also monitor the number of clusters and the size of the largest cluster as a function of time during the production phase, and find no overall drift that would suggest the system is undergoing macroscopic phase separation.

\begin{figure}\bigskip
\centerline{\includegraphics[scale=0.37]{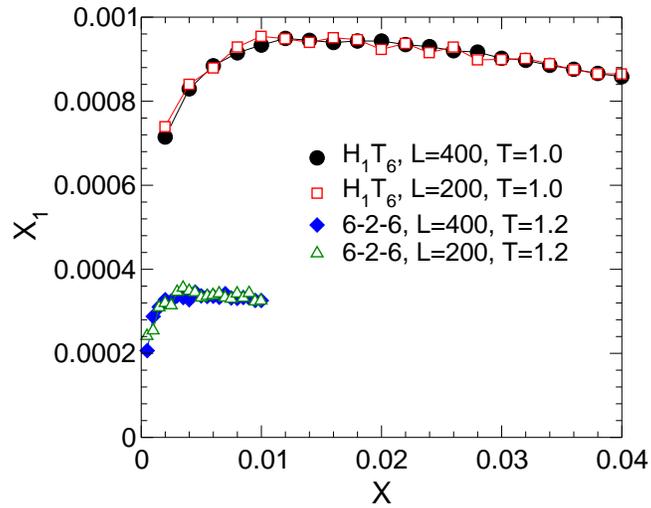}}
\caption{$X_1$ as a function of $X$ for the H$_1$T$_6$ and 6-2-6 systems for both $L=200$ and $400$.
}
\label{c3}
\end{figure}

(iv) Finally we test for finite-size effects in two ways.  First, we evaluate $X_1$ as a function of $X$ for systems with both $L=200$ and $L=400$ (Fig.~\ref{c3}).  The curves for these two system sizes coincide within statistical error, indicating that a system of size $L=200$ is large enough to be free of significant finite-size effects.  Second, we show in Fig.~\ref{c2} $P(n)$ at the lowest $T$ for both $L=200$ and $L=400$.  Again, the curves coincide, supporting the absence of finite-size effects for the $L=200$ system.  This second test is also consistent with the absence of macroscopic phase separation:  For a finite system undergoing phase separation,
the $n$ value for the peak in $P(n)$ will increase as $L$ increases, since larger clusters are possible in a larger system~\cite{Lisal:2002p6038}.  However, for a micelle-forming system, $P(n)$ is independent of system size, consistent with the behavior observed here.

\subsection{Evaluating the CMC}

We determine $X_{\rm CMC}$ from a plot of $X_1$ versus $X$.  
Our curves for $X_1$ versus $X$ pass through a maximum value $X_1^{\rm max}$.
To define the CMC we use the same definition as in Refs.~\cite{Tanford:1980p7160,Zaldivar:2003p6348}, in which $X_{\rm CMC}$ is defined as the value of $X$ at which the curves $X_1=X_1^{\rm max}$ and $X_1=X$ intersect.  
That is, $X_{\rm CMC}=X_1^{\rm max}$.

\begin{figure}\bigskip\bigskip
\centerline{\includegraphics[scale=0.37]{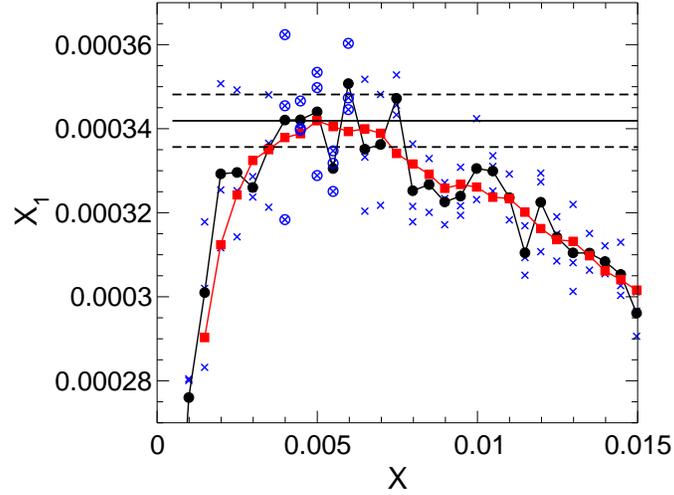}}
\caption{$X_1$ as a function of $X$ for the 6-2-6 system for $T=1.2$.  We compute the value of $X_{\rm CMC}$ (solid horizontal line) as the maximum value of the smoothed data set for $X_1$ as a function of $X$ (red squares), as described in the text.  Also shown are the values of $X_1$ obtained from individual runs (blues crosses) as well as the average of $X_1$ over the three runs conducted at each $X$ (black circles).  The error in $X_{\rm CMC}$ (horizontal dashed lines) is twice the standard deviation of the mean of the 15 data points (circles with crosses) occurring in the vicinity of the maximum value of $X_1$.
}\label{error}\end{figure}

To compute $X_1^{\rm max}$, we smooth the data by averaging $X_1$ over successive groups of five data points in the interval $X-0.001\ge X \le X+0.001$, for each value of $X$ in the vicinity of the maximum.
We estimate $X_1^{\rm max}=X_{\rm CMC}$
as the largest value of the smoothed $X_1$ data; see Fig.~\ref{error}.

We estimate the statistical error of $X_{\rm CMC}$ from the scatter in the independent runs used to compute $X_1$ in the vicinity of the maximum (Fig.~\ref{error}).  Let $X_{\rm max}$ be the value of $X$ at which the maximum occurs in the smoothed data for $X_1$.  
For each $X$ we have three independent runs, and therefore three independent evaluations of $X_1$ at each $X$ near $X_{\rm max}$.
Further, the variation of $X_1$ in the vicinity of $X_{\rm max}$ is weak (because it's a maximum), and so we assume the $X_1$ measurements for the five values of $X$ in the interval $X_{\rm max}-0.001\ge X_{\rm max} \le X_{\rm max}+0.001$ are all estimates of $X_{\rm CMC}$.  This provides us with a set of 15 independent estimates of $X_{\rm CMC}$, for which we compute the standard deviation $\sigma$.  The error in $X_{\rm CMC}$ is taken as $\pm 2\sigma/\sqrt{15}$, that is, twice the standard deviation of the mean for a sample size of 15.  As shown in the following, we find that the computed error for $X_{\rm CMC}$ is smaller than the symbol size used in our plots.

\begin{figure}\bigskip
\centerline{\includegraphics[scale=0.37]{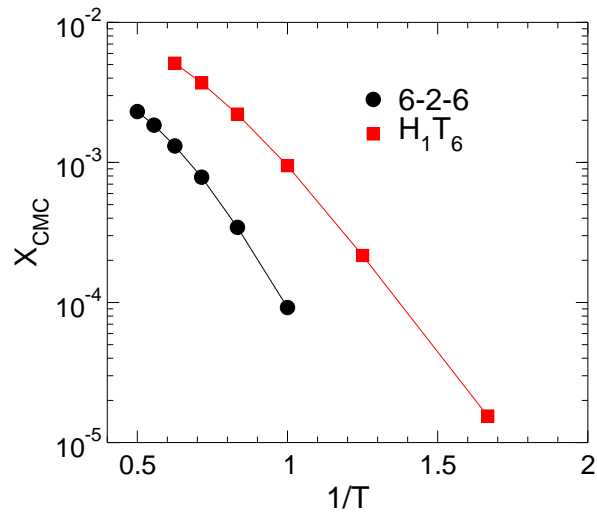}}
\caption{Arrhenius plot of the variation of $X_{\rm CMC}$ with $T$, for the H$_1$T$_6$ and 6-2-6 systems.
}\label{tdep}\end{figure}

In Fig.~\ref{tdep} we show the $T$ dependence of $X_{\rm CMC}$ 
for both the H$_1$T$_6$ and 6-2-6 cases.   The curves have the expected form, in that $X_{\rm CMC}$ decreases with $T$.  At the lowest $T$, in the micelle-forming regime, $X_{\rm CMC}$ approaches an Arrhenius dependence on $T$, as also found in previous simulations for micelle-forming systems~\cite{Panagiotopoulos:2002p6025,Davis:2008p5959}.  Consistent with experiments, the CMC of the gemini 6-2-6 system is up to an order of magnitude lower than the corresponding single-tailed surfactant H$_1$T$_6$~\cite{Zana:2002p6378}.

\section{Results}

We now present our results for the CMC of all distinct amphiphile architectures containing two head units, for fixed $N=12$ and $16$.  We carry out simulations of all these architectures using the computational protocol described above.  All simulations are conducted for $T=1.2$.  As we will see below, at this $T$ we find that all the architectures give systems in which some degree of micelle-like aggregation is occurring.

\subsection{Influence of the number of spacer units}

\begin{figure}
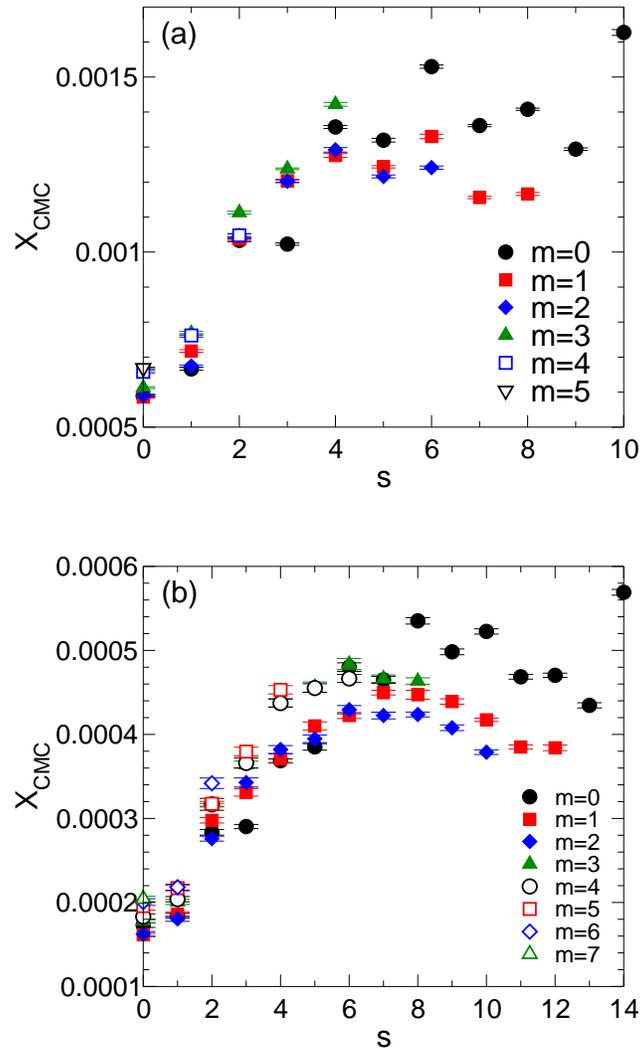
\bigskip
\centerline{\includegraphics[scale=0.37]{fig-9a.eps}}
\bigskip\bigskip
\centerline{\includegraphics[scale=0.37]{fig-9b.eps}}
\caption{$X_{\rm CMC}$ for all distinct combinations of $m$ and $s$ for (a) $N=12$ and (b) $N=16$.  For all points, $T=1.2$.
}\label{cmc-s}\end{figure}

Our results for the CMC of all the distinct architectures are shown in Fig.~\ref{cmc-s}.  In Fig.~\ref{cmc-s} we plot $X_{\rm CMC}$ 
as a function of $s$, and group points corresponding to constant values of $m$ by using the same symbol type.  Note in all cases that the statistical error in $X_{\rm CMC}$ is smaller than the symbol size.

\begin{figure}\bigskip
\centerline{\includegraphics[scale=0.37]{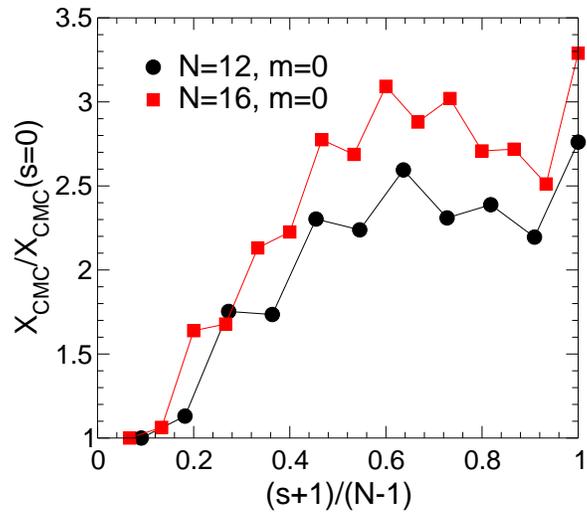}}
\caption{$X_{\rm CMC}$ as a function of $s$ for $m=0$, for both $N=12$ and $N=16$.  The scaling of the vertical axis facilitates comparison of the effect on the CMC relative to the $s=0$ case.  The horizontal axis quantifies the path length along the chain between the heads ($s+1$) relative to the total path length of the chain ($N-1$).
}\label{cmc-s-scaled}\end{figure}

For $s<N/2$, the CMC increases with $s$.  The scale of the increase is a factor of approximately 2.5 relative to the lowest CMC at $s=0$; see Fig.~\ref{cmc-s-scaled}.  
While the data are more scattered for $s>N/2$, there is a trend for the CMC values to pass through a maximum and begin to decrease as $s$ increases further.  This is most evident in the $m=1$ and $m=2$ curves for $N=16$ [Fig.~\ref{cmc-s}(b)].  
This behavior, where the CMC initially increases with $s$ and then passes through a maximum, is consistent with several experimental studies~\cite{ 
Zana:1991p6462,
Borse:2005p6376,
Sikiric:2005p6375,
VanBiesen:2007p6377}, 
as well as earlier computer simulations~\cite{
Maiti:1998p6401}.
Most of the experimental work reporting this phenomenon has been conducted for symmetric gemini surfactants having a fixed tail length where the size of the spacer segment is progressively increased; hence the total length $N$ of the amphiphile chain is increasing.  It is interesting to note that the same behavior occurs in our system, where $N$ is fixed regardless of how $s$ changes, and where both symmetric and asymmetric cases are considered.

We also note the ``saw-tooth" variation of $X_{\rm CMC}$ as a function of $s$, which is most apparent for small $m$ (Fig.~\ref{cmc-s-scaled}).  Since our statistical errors are smaller that the symbol size, these variations must be a systematic effect, and are most likely due to the lattice discretization inherent in the model.

The most notable exception to the trend in the CMC as a function of $s$ occurs for the boloform amphiphiles, which have the largest value of $s=N-2$ and a head unit located exactly at each end of the chain.  $X_{\rm CMC}$ for the boloform amphiphiles is the largest of all the architectures for both $N=12$ and $16$.  The behavior of $X_{\rm CMC}$ at large $s$ is therefore complex, and we return to this issue in Section III.C below.

\begin{figure}\bigskip
\centerline{\includegraphics[scale=0.37]{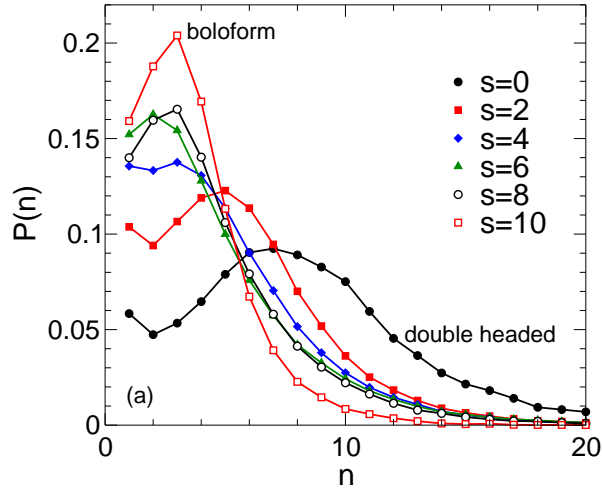}}
\bigskip\bigskip
\centerline{\includegraphics[scale=0.37]{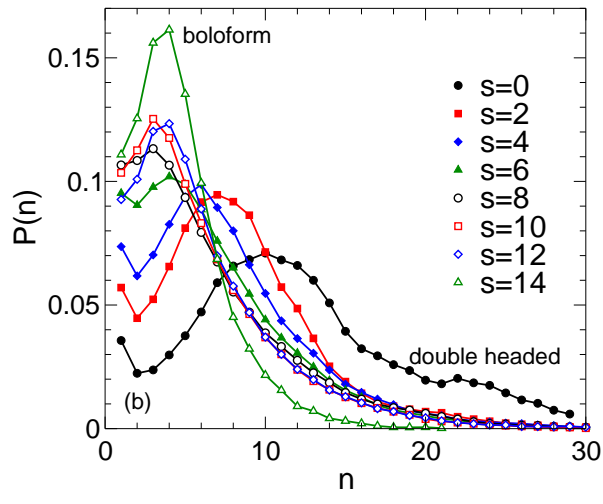}}
\caption{Influence of $s$ on $P(n)$ for $T=1.2$.  For all curves $m=0$, i.e. one head group is fixed at one end of the chain, while the second head group moves along the chain as $s$ increases.  $s=0$ corresponds to a double-headed amphiphile, while the largest value of $s$ corresponds to a boloform amphiphile.  In (a) $N=12$ and  $X=0.01$; in (b) $N=16$ and $X=0.005$.
}\label{Pn-s}\end{figure}

\begin{figure}
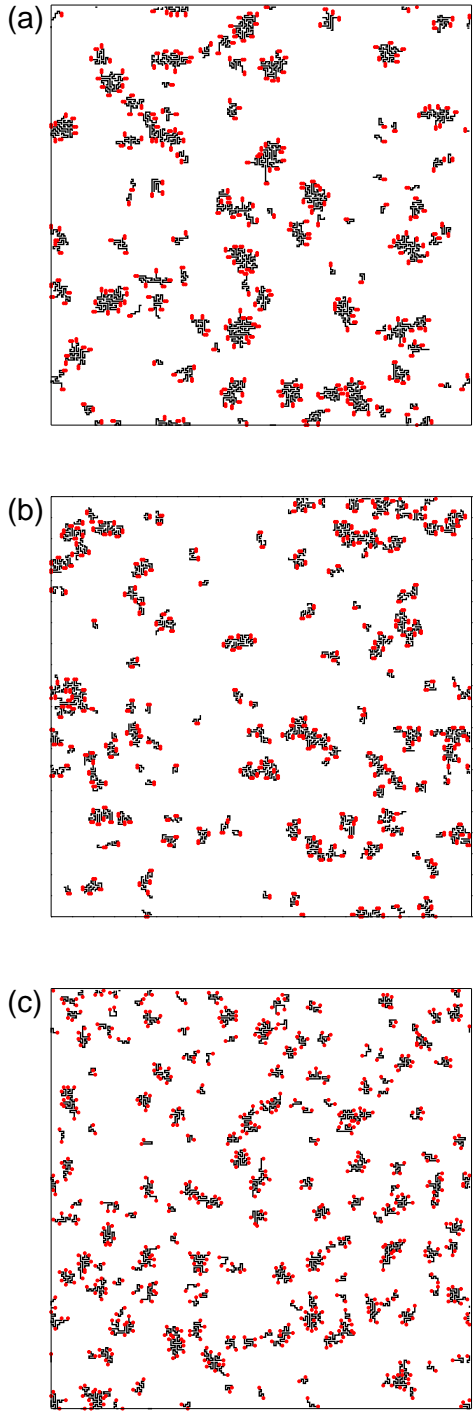
\bigskip
\centerline{\includegraphics[scale=0.37]{fig-12a.eps}}
\bigskip\bigskip
\centerline{\includegraphics[scale=0.37]{fig-12b.eps}}
\bigskip\bigskip
\centerline{\includegraphics[scale=0.37]{fig-12c.eps}}
\caption{Snapshots of the system for $L=200$, $N=12$, $X=0.01$ and $T=1.2$ for (a) $m=0$ and $s=0$ (double-headed case); (b) $m=5$ and $s=0$ (symmetric gemini case); and (c) $m=0$ and $s=10$ (boloform case).  Tail units are rendered as black line segments, and head units are red dots.
}\label{pic-s}\end{figure}

In Fig.~\ref{Pn-s} we quantify the changes in the morphology of the amphiphile aggregates by plotting $P(n)$ for several values of $s$, all at constant $m=0$.  This series corresponds to fixing one head unit on one end of the chain, and moving the second head unit along the chain.  These $P(n)$ curves therefore progress from the double-headed architecture ($s=0$) through to the boloform case ($s=N-2$).  For both $N=12$ and $16$ we find that the double-headed architecture produces the largest aggregates, as characterized by the value of $n$ at the maximum in $P(n)$; as well as the widest distribution of aggregate sizes, as characterized by the width of the peak.  A snapshot of the system for the double-headed case is given in Fig.~\ref{pic-s}(a).

As $s$ increases up to approximately $s=N/2$, the size of the aggregates decreases, and the peak becomes less distinct, indicating that the micelle-like character of the amphiphile aggregates is degrading.  This regime coincides with the range of $s$ in which $X_{\rm CMC}$ is increasing with $s$.  For $N/2<s<N-2$, the distinctness of the peak in $P(n)$ recovers somewhat, though the aggregates remain small compared to $s=0$.  This modest recovery of the micelle-like morphology is consistent with the trend for $X_{\rm CMC}$ to decrease over this range of $s$  (with the exception of the boloform case noted above).  The $P(n)$ curve for the boloform amphiphiles ($s=N-2$) has the sharpest peak of all the curves, but the typical aggregate size remains small.  These observations are confirmed by examining a snapshot of the system for the boloform case [Fig.~\ref{pic-s}(c)] as compared to the system with double-headed amphiphiles [Fig.~\ref{pic-s}(a)].  

\begin{figure}
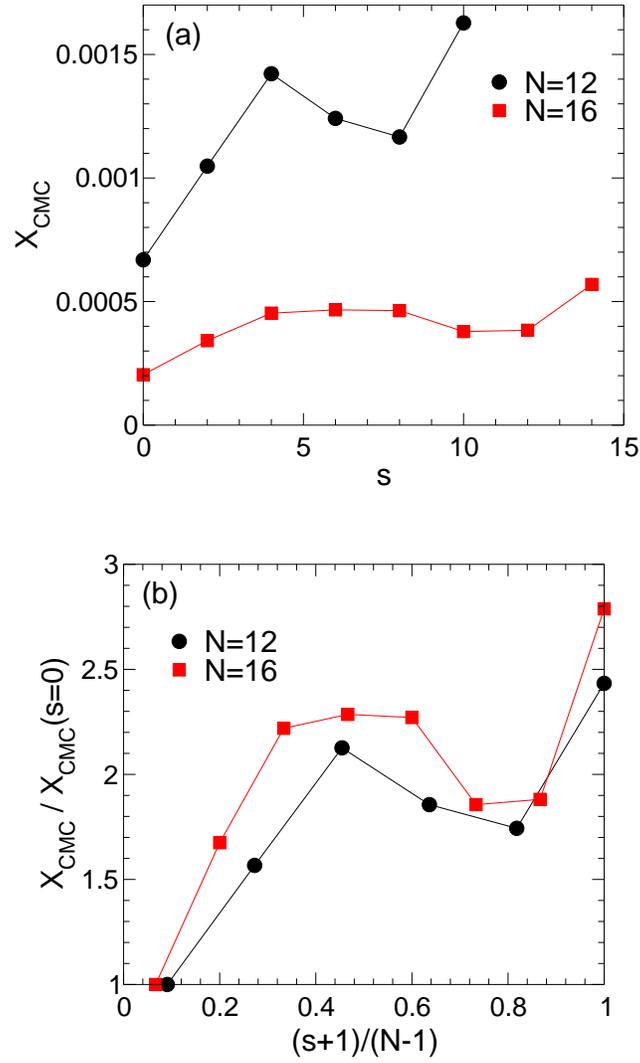
\bigskip
\centerline{\includegraphics[scale=0.37]{fig-13a.eps}}
\bigskip\bigskip
\centerline{\includegraphics[scale=0.37]{fig-13b.eps}}
\caption{$X_{\rm CMC}$ as a function of $s$ for symmetric amphiphiles.  In (b), the axes are scaled as in Fig.~\ref{cmc-s-scaled} to facilitate comparison of the results for different chain lengths.  Note that all points correspond to symmetric gemini surfactants except for the largest $s$ value, which corresponds to a boloform amphiphile.
}\label{cmc-sym}\end{figure}

\begin{figure}\bigskip
\centerline{\includegraphics[scale=0.37]{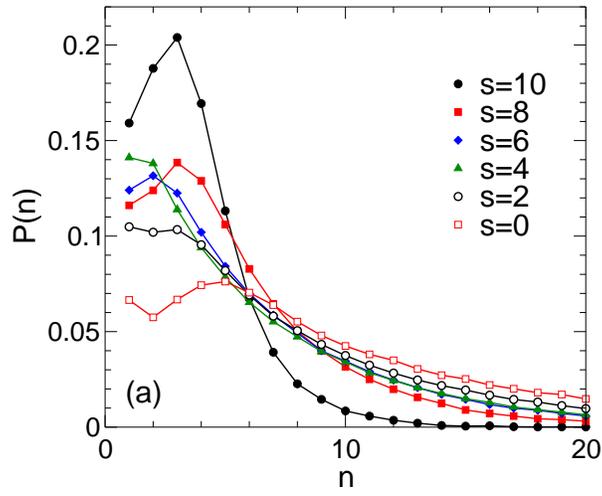}}
\bigskip\bigskip
\centerline{\includegraphics[scale=0.37]{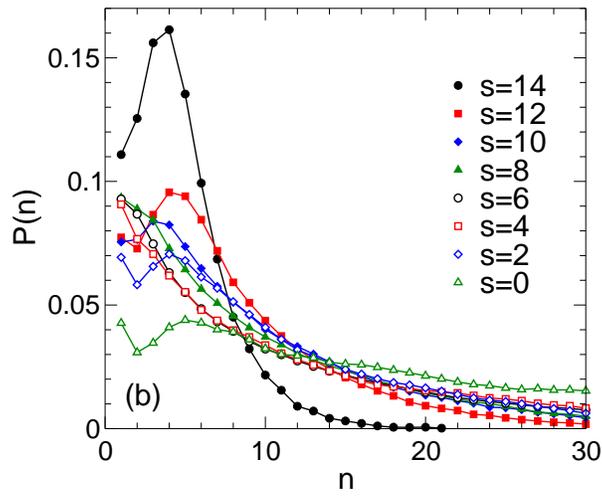}}
\caption{Influence of $s$ on $P(n)$ for symmetric amphiphiles.  In (a) $N=12$ and $X=0.01$; in (b) $N=16$ and $X=0.005$.  
}\label{Pn-sym}\end{figure}

As shown in Figs.~\ref{cmc-sym} and \ref{Pn-sym}, the same general trends for the influence of $s$ on the CMC and aggregate morphology are observed if we restrict our attention solely to symmetric amphiphiles.  These symmetric architectures include all the distinct gemini surfactants having equal tail lengths, as well as the bolofom architecture.  For $s<N/2$, increasing the number of spacer units, at the expense of the length of the tails, increases the CMC, and degrades the formation of aggregates with a distinct size.  We note however that the typical size of the aggregates varies less than in the progression shown in Fig.~\ref{Pn-s}.  As $s$ increases beyond $N/2$ there is initially some recovery of distinct micelle-like aggregates [observed as a reappearance of the peak in $P(n)$], and a lowering of the CMC.  However, again the boloform case is a special one, in which the CMC is highest, while the aggregates have a very distinct but small size.

Experimentally, it has been observed for symmetric gemini surfactants that the typical size of aggregates decreases as $s$ increases~\cite{
Zana:2002p6378,
Dreja:1999p7163,
Borse:2005p6376}.  These experiments were conducted using $m$-$s$-$m$ architectures in which $m$ was held constant and $s$ increased, and hence the total amphiphile length was increased.  Our results show that it is also possible to realize similar changes in aggregate morphology by varying $s$ at constant $N$, although the effect in the case of symmetric amphiphiles is relatively weaker than for the asymmetric architectures.

\subsection{Influence of amphiphile symmetry}

Implicit in the results shown in Fig.~\ref{cmc-s} is the finding that there is little dependence of $X_{\rm CMC}$ on $m$, especially for small values of $s$.  Since $m$ determines the placement of the first head unit along the chain, it controls the symmetry of the architecture for a given value of $s$.  However, the degree of symmetry is not immediately apparent from a given $(m,s)$ pair.  Accordingly, we define an ``asymmetry index" by considering the distance $D=(N-s-2m-2)/2$, which for a straight chain is the distance from the midpoint between the two head units, to the midpoint of the chain.  For a symmetric gemini or boloform surfactant, $D=0$.  The maximally asymmetric case is the double-headed amphiphile, for which $D=D_{\rm max}=(N-2)/2$.  We thus define our asymmetry index as $a=D/D_{\rm max}$, so that the asymmetry of the architecture is measured relative to the maximum possible asymmetry.

\begin{figure}
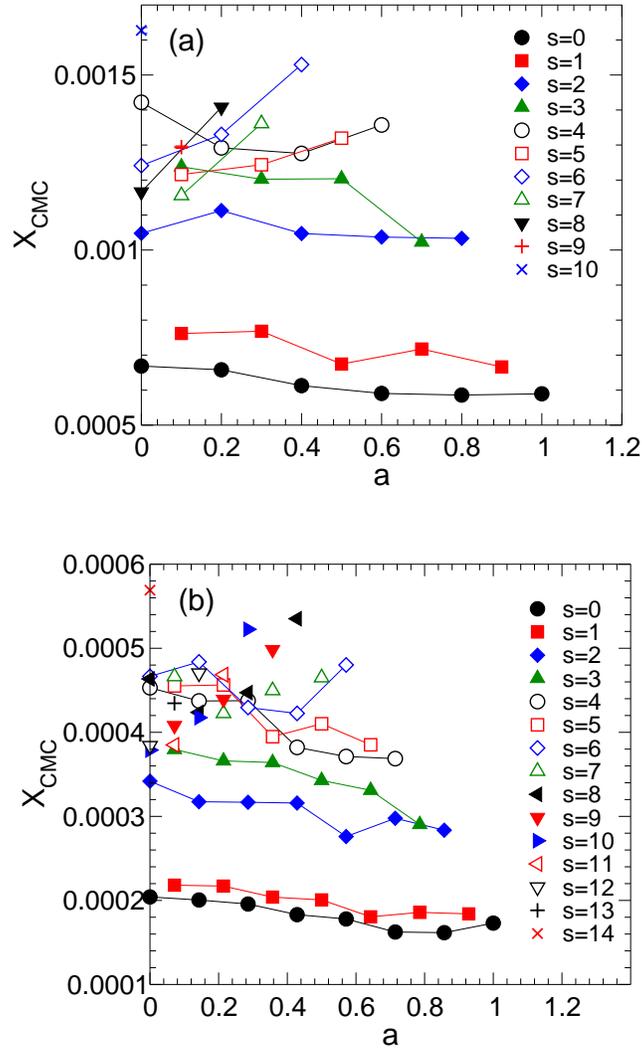
\bigskip
\centerline{\includegraphics[scale=0.37]{fig-15a.eps}}
\bigskip\bigskip
\centerline{\includegraphics[scale=0.37]{fig-15b.eps}}
\caption{Influence of the amphiphile asymmetry $a$ on $X_{\rm CMC}$ at $T=1.2$, for (a) $N=12$ and (b) $N=16$.  
The asymmetry index is defined as $a=(N-s-2m-2)/(N-2)$.
For example, a double-headed amphiphile has $a=1$, while a symmetric gemini surfactant or a boloform amphiphile both have $a=0$.  
}\label{cmc-a}\end{figure}

Fig.~\ref{cmc-a} displays the same data for $X_{\rm CMC}$ as shown in Fig.~\ref{cmc-s}, but reparametrized so that it is plotted as a function of $a$, with groups of points having the same $s$ value sharing the same symbol.  Fig.~\ref{cmc-a} shows that there is only a weak dependence of the CMC on the symmetry of the amphiphile architecture.  The dependence is weakest for small $s$ values.  For large $s$ there is more variation with $a$, but little in the way of a pattern from one $s$ value to the next.  At fixed $s$, the variation of $X_{\rm CMC}$ never exceeds a factor $0.2$.  This is small compared to the relative changes found as a function of $s$, which are a factor of approximately $2.5$.

\begin{figure}
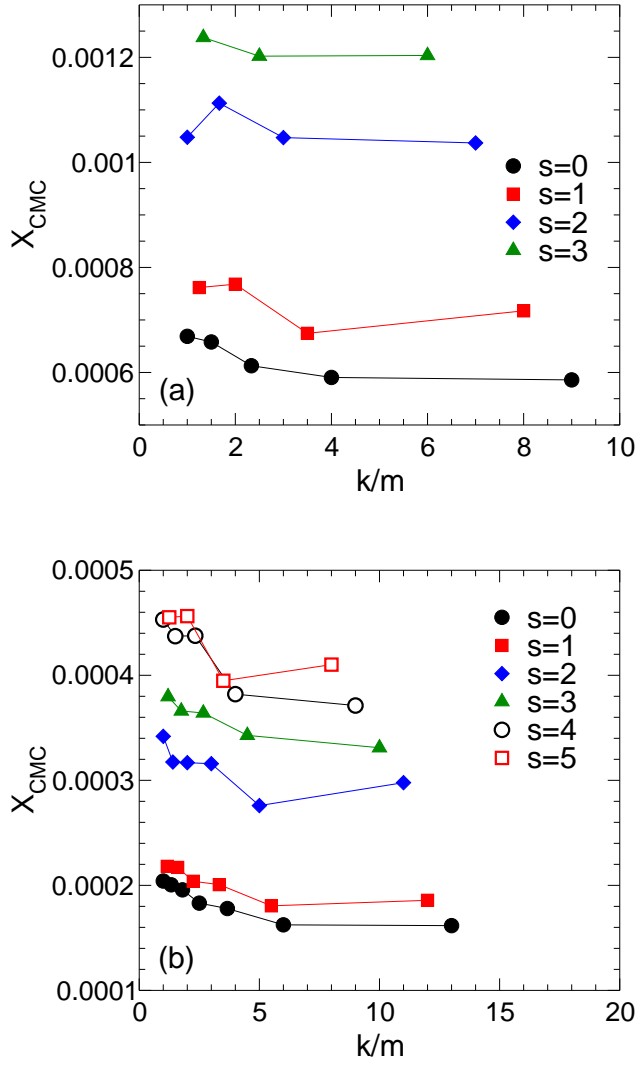
\bigskip
\centerline{\includegraphics[scale=0.37]{fig-16a.eps}}
\bigskip\bigskip
\centerline{\includegraphics[scale=0.37]{fig-16b.eps}}
\caption{Influence of the asymmetry of gemini amphiphiles on $X_{\rm CMC}$ at $T=1.2$, for (a) $N=12$ and (b) $N=16$.  Here the
asymmetry is quantified by $k/m$, where $k$ is the number of tail units in the longer of the two tails, while $m$ is the number in the shorter tail.  
For example, a symmetric gemini surfactant has $k/m=1$, while the most asymmetric gemini surfactant for $N=16$ and $s=0$ has $k/m=13$.  Note that amphiphiles with $m=0$ are excluded from this plot since in this case $k/m$ diverges.
}\label{cmc-km}\end{figure}

As noted in the introduction, Thomas and coworkers~\cite{Bai:2002p6476,Wang:2003p6498} investigated $m$-$s$-$k$ architectures of dissymmetric gemini surfactants for values of $k/m$ from 1 to 3.  They found that the CMC decreased by approximately 35\% over this range.  To facilitate a direct comparison, in Fig.~\ref{cmc-km} we replot our CMC data versus $k/m$ for several values of $s$, for architectures having $m>0$.  As in Refs~\cite{Bai:2002p6476,Wang:2003p6498} we also find a decrease in the CMC for most value of $s$, although the effect is somewhat weaker, despite the fact that our data extends up to $m/k=13$.  

\begin{figure}\bigskip
\centerline{\includegraphics[scale=0.37]{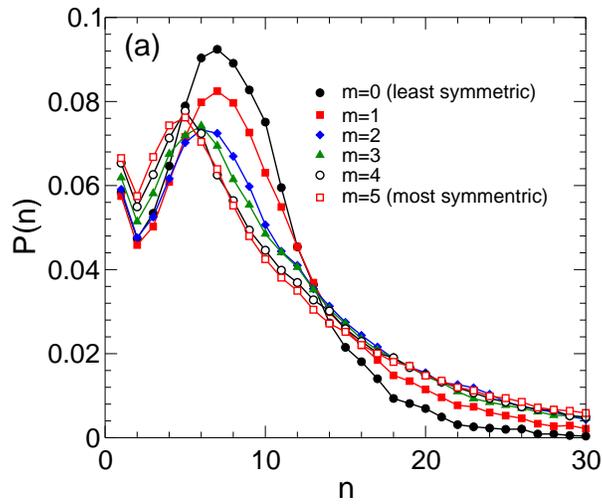}}
\bigskip\bigskip
\centerline{\includegraphics[scale=0.37]{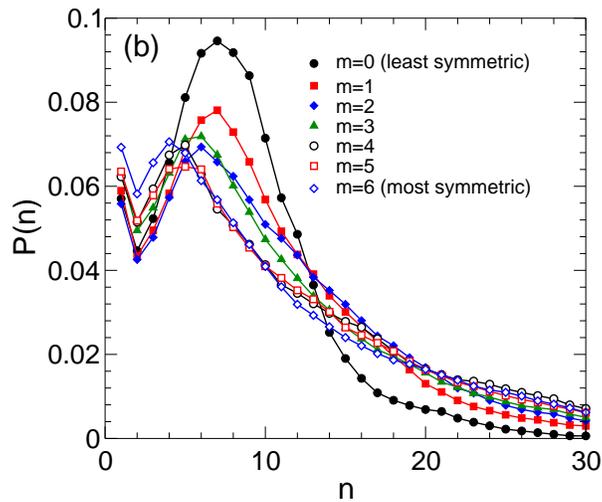}}
\caption{Influence of amphiphile symmetry on $P(n)$.  In (a) $N=12$, $s=0$ and $X=0.01$; in (b) $N=16$, $s=2$ and $X=0.005$.  
In each panel, the smallest value of $m$ is the least symmetric, while the largest value corresponds to a symmetric gemini surfactant.
}\label{Pn-a}\end{figure}

The behavior of $P(n)$ as a function of asymmetry at fixed $s$ is consistent with the variation of the CMC noted above.  As shown in Fig.~\ref{Pn-a}, while there are changes in the height and position of the peak in $P(n)$, these are smaller changes than those observed due to changes in $s$ (Fig.~\ref{Pn-s}).  For the small values of $s$ depicted in Fig.~\ref{Pn-a} the peak is present for all values of $m$, suggesting that the degree of aggregate formation is not qualitatively disrupted as the symmetry of the amphiphile architecture is varied.  This is in contrast to the suppression of the peak in $P(n)$ that we observe for intermediate values of $s$ in Figs.~\ref{Pn-s} and \ref{Pn-sym}.  While the change in the typical size of the aggregates is relatively smaller in Fig.~\ref{Pn-a}, the direction is the same as found in experiments: Ref.~\cite{Wang:2003p6498} reports that the size of aggregates increases as the degree of asymmetry increases, as is found here.

\subsection{Behavior of boloform amphiphiles}

As noted above, the behavior of the CMC for the bolofom architectures studied here does not follow the trend established at smaller values of $s$ (Fig.~\ref{cmc-s}).  Our results for the CMC of the boloform architecture is also inconsistent with experimental results.  Based on a comparison of the experimental value of the CMC for an 8-6-8 gemini surfactant~\cite{Frindi:1994p7162} with that of a boloform amphiphile of approximately equal overall length~\cite{Hirata:1995p7164}, we would conclude from published data that the boloform architecture should have a CMC that is smaller by almost an order of magnitude.  This is clearly at odds with our result that the boloform amphiphile has the highest CMC of any architecture studied here.

This discrepancy is all the more perplexing because the morphology of the aggregates formed by the boloform amphiphiles in our simulations is consistent with experiments.  Boloform surfactants have long been recognized for producing unusually small micellar aggregates in comparison to other amphiphiles~\cite{
Yiv:1976p7141,
Ikeda:1989p6383,
Moroi:1992p6386}.  This is also true here.  The typical sizes of our boloform aggregates are the smallest of all the architectures we examine (Figs.~\ref{Pn-s}).  

Given the general agreement between our results and experimental trends for most other architectures, the anomalously high values we find for the CMC of the boloform case is most likely due to the idealized nature of our model.  In particular, our use of a 2D lattice may be qualitatively changing the behavior of the boloform case.

\section{Conclusions}

Our model amphiphiles are highly idealized, and real surfactants are obviously more complex.  For example, we have ignored the effect of chain stiffness or restrictions in bending angles that exist, especially in the bonds that connect unlike segments of the structure, such as the bonds linking head groups to hydrocarbon chains.  Also, many of the real surfactants that motivate our work are not truly linear chains.  For example, in many gemini surfactants the spacer does not connect directly to the head groups, but rather at some other point along the tails.  Restricting the dimensions of the lattice to 2D may also introduce serious distortions into the results as compared to 3D, as we suspect may be the case for the behavior of our boloform amphiphiles.  That said, we note that the case of micelle formation in quasi-2D systems such as on surfaces and in thin films may be interesting in its own right~\cite{Israelachvili:1994p7137}.

Nonetheless, the qualitative trends in most of our results are in line with experimental findings.  Our main result is to show for a complete family of two-headed linear amphiphiles, all of the same overall length and analyzed in the same way, that both the size of the spacer and the amphiphile asymmetry play a role in determining the CMC and the aggregate morphology;  however, the influence of the spacer size is about an order of magnitude more than that due to asymmetry.  This result is in line with the findings reported in the experimental literature on symmetric and dissymmetric gemini surfactants.  The fact that we observe a relative influence of spacer length and asymmetry similar to that found in experiments, using such a simple and highly idealized model, strongly suggests the fundamental geometric origin of this relationship.

Finally, our work illustrates that an exhaustive examination of amphiphile architectures (in the spirit of ``combinatorial chemistry") is computationally possible using models of the kind studied here.  Our results demonstrate that this approach can help elucidate and quantify structure-property relationships that otherwise must be deduced from an analysis of multiple studies, often conducted using different measurement techniques and chemically distinct species.

\section{Acknowledgements}

We thank ACEnet for providing computational resources.  Funding was provided by AIF, NSERC and the CRC program.

%\bibliographystyle{prsty}
%\bibliography{surf}

\end{document}